\documentclass[english,pre,twocolumn,showkeys,showpacs,floatfix]{revtex4}
\usepackage[T1]{fontenc} \usepackage[latin1]{inputenc}
\usepackage{babel} \usepackage{epsfig}
\usepackage{amsmath}\newcommand{\pde}[1]{\frac{\partial}{\partial
    {#1}}}
\usepackage[percent]{overpic}
\newcommand{\pdz}[1]{\frac{\partial^2}{\partial{#1}^2}}
\newcommand{\tde}[1]{\frac{d}{d {#1}}}

\newcommand{\pd}[2]{\frac{\partial^{#1}}{\partial{#2}^{#1}}}
\newcommand{\LT}{{\mathbf M}}
\newcommand{\JJ}{{\cal{J}}}
\newcommand{\JJJ}{{\mathbf{J}}}
\newcommand{\PP}{{\cal{P}}}

\newcommand{\mean}[1]{\left<{#1}\right>}
\newcommand{\pv}{{\mathbf{p}}}
\newcommand{\qv}{{\mathbf{q}}}

\begin{document}

\title{Phase velocity and phase diffusion in periodically driven discrete
  state systems} \author{T.~Prager and L.~Schimansky-Geier}
\affiliation{Institute of Physics, Humboldt-University of Berlin,
  Newtonstr. 15, 12489 Berlin, Germany } \date{\today}

\begin{abstract}
  We develop a theory to calculate the effective phase diffusion
  coefficient and the mean phase velocity in periodically driven
  stochastic models with two discrete states. This theory is
  applied to a dichotomically driven Markovian two state system
  Explicit expressions for the mean phase velocity, the effective
  phase diffusion coefficient and the P\'eclet number are analytically
  calculated. The latter shows as a measure of phase-coherence forced
  synchronization of the stochastic system with respect to the
  periodic driving.  In a second step the theory is applied to a non
  Markovian two state model modeling excitable systems. The results
  prove again stochastic synchronization to the periodic driving and
  are in good agreement with simulations of a stochastic
  FitzHugh-Nagumo system.
\end{abstract}

 \maketitle

\section{Introduction}
Stochastic resonance as a phenomenon of noise enhanced order in
periodically driven stochastic systems has attracted considerable
interest until today \cite{benzi,cnicolis,gammaitoni,uspekhi}. A
common approach to quantify this effect are spectral based measures like the spectral power amplification
and the signal to noise ratio. On the other hand stochastic resonance can also be understood as a synchronization
process between the input and the response of the system
\cite{gammaitoni,ani_book}. This interpretation achieves importance
especially if dealing with larger amplitudes of the driving signal.
Then analytical descriptions have to go beyond linear response theory.

In general two principal approaches were introduced in the past to
describe the synchronization of a stochastic system by an external
driving. The first one bases on the consideration of escape time
densities to leave certain states of the dynamical system \cite{moss}.
A periodic driving modulates these densities and they exhibit maxima
at times which corresponds to time scales of the external drive.
``Bona fide'' resonances were investigated analytically, numerically
simulated and and experimentally verified, especially for symmetric
bistable situations \cite{santucci,choi,giacomelli,talkner}.

The second approach goes back to Stratonovich who looked at
synchronization of nonlinear oscillators by periodic driving in the
presence of noise \cite{strato}. For this purpose one adopts a phase
to the nonlinear oscillators and defines statistical properties of the
stochastically behaving phase. If the mean phase velocity agrees with
the frequency of the driving and at the same time the phase diffusion
coefficient is small then there exist in average a fixed phase relations
between the driving and the output of the system.

This picture was recently transfered to models of stochastic resonance
which are nonlinear but non oscillating. It was possible to prescribe a
phase to overdamped bistable as well as to excitable systems which
monotonously increases in time \cite{neiman,collins,lindner_rep}. Its
mean velocities and effective phase diffusion constant were used to
quantify synchronization between the output and the driving input.
Likewise as in stochastic resonance synchronization appears at an
optimal choice of the noise intensity since the level of noise
determines the characteristic times of the stochastic system.  

As result one finds plateaus of the mean frequencies of the output at
values which correspond to the driving frequency or multiples of it
\cite{ani_book,shulgin,longtin2,zhou,laser}.  These plateaus are accompanied
with low phase diffusion coefficients indicating a synchronization in
average.  As measure of synchronization one uses the duration of
locking epoches or a P\'eclet-number which is the ratio between the
phase velocity and phase diffusion coefficient
\cite{freund,freund2,lai}.

For bistable stochastic systems a discrete state modeling has been
proven very successful in the past \cite{wiesenfeld,loefstedt}.  It is
based on a separation of time scales between the fast relaxation into
the metastable states and the transition between these states, which
happens on a slower time scale and build up a Markovian discrete
dynamics \cite{talkner3}.

Also models of excitable behavior \cite{longtin,wiesenfeld2} can be
mapped on two or  three state dynamics
\cite{lindner,prager1,prager2}. These discrete state models still set
up a renewal process \cite{cox}.
However in difference to bistable systems
they include non-exponentially distributed waiting time densities and
are thus non Markovian.

These discrete state systems will be endowed with a discrete phase which is
introduced in Section \ref{sec.2}. As will be shown in our paper both 
the Markovian and the non Markovian model exhibit phase synchronization with respect to the periodic
driving for optimal noise levels. We will quantify this effect by the
mean phase velocity, phase diffusion coefficient and the P\'eclet
number. An unique approach to calculate these quantities in driven
renewal models with two states will be presented in Section
\ref{sec.3}. This approach is based on an envelope description of the
phase \cite{freund2,harms}.  

Section \ref{bistable} applies the theory to bistable systems where
Markovian rules were assumed for the transition between the discrete
states.  First results of this system with dichotomic periodic inputs
were derived earlier in \cite{freund}.  These results were recently
improved in \cite{talkner2,casado2} which agrees with our findings in
case of Markovian dynamics.

Section \ref{excitable} is devoted to a non Markovian two state system
which models excitable behavior. Integral equations for the phase
velocity and phase diffusion coefficient have to be numerically
solved. Results of these computations show good quantitative agreement
with numeric simulations of a stochastic periodically driven
FitzHugh-Nagumo system.

\section{Two state models and phase}
\label{sec.2}
Consider a periodically driven stochastic two state system described
by the probabilities $\pv(t)=[p_1(t),p_2(t)]$ to be in state 1 or 2
respectively at time $t$.

In generally these dynamics can be expressed in terms of the  flux
operators $\JJ^{i\to j}_{t}$ by
\begin{eqnarray}\label{generaleqp1}
\dot p_1&=&\JJ_t^{2\to1}[\pv(\cdot)]-\JJ_t^{1\to2}[\pv(\cdot)]\\
\dot p_2&=&\JJ_t^{1\to2}[\pv(\cdot)]-\JJ_t^{2\to1}[\pv(\cdot)]\label{generaleqp2}
\end{eqnarray}
The linear flux operators, which express  the probability flux from state
$i$ to state $j$ in terms of the occupation probabilities
depend  explicitly on
time $t$ in a periodic way due to the periodical
driving with period $T=2\pi/\Omega$,
\begin{eqnarray}
\JJ^{i\to j}_{t}=\JJ^{i\to j}_{t+T}.\label{opperiodicity}
\end{eqnarray}
In the Markovian case these operators are local in time, i.e.  multiplication
operators, 
\begin{eqnarray*}
\JJ^{i\to j}_{t}[\pv(\cdot)](t)=\gamma_i(t) p_i(t). 
\end{eqnarray*}
The well known two state model for bistable systems \cite{wiesenfeld}
which will be considered in more detail in section \ref{bistable} is
of this type.
In the non Markovian case  the action of the flux operators
$\JJ^{i\to j}_{t}$ on the probabilities $p_1$ and $p_2$ is non-local in time,
i.e. the $\JJ^{i\to j}_{t}$ are integral operators.
One example of this type is the discrete state model for excitable systems
\cite{prager1,prager2}, whose flux operators are given by
\begin{eqnarray*}
  \JJ^{1\to 2}_{t}[\pv(\cdot)](t)&=&\int_{t_0}^t d\tau w(t-\tau)\gamma(\tau)p_2(\tau)\\
  \JJ^{2\to 1}_{t}[\pv(\cdot)](t)&=&\gamma(t)p_2(t)
\end{eqnarray*}
Note that in this case the flux operator depend explicitly on the
initial time $t_0$ which breaks its periodicity eq.
(\ref{opperiodicity}). However in the asymptotic case $t_0\to-\infty$
this periodicity is restored.  This model will be considered in
section \ref{excitable}.

Next we endow this system with a phase $\phi(t)$.
Our goal is to evaluate the mean phase velocity 
\begin{eqnarray}\label{definitionomega0}
  \bar \omega:=\lim_{t\to\infty} \frac{\mean{\phi(t)}}{t}
\end{eqnarray}
as well as the effective phase diffusion constant
\begin{eqnarray}\label{definitionD0}
  \bar D_\text{eff}:=\lim_{t\to\infty} \frac{\mean{\phi^2(t)}-\mean{\phi(t)}^2}{2 t}. 
\end{eqnarray}
These quantities are independent of
the exact definition of phase, as long as the phase
increases by $2\pi$ within a one cycle $1\to 2\to 1$ of the system.
For the sake of notational and computational 
convenience we consider a phase, which increases by
$2\pi$ each time the system enters state 1. 
Then the probabilities   $\pv_k=[p_{1,k},p_{2,k}]$ 
to be in state 1 or 2 respectively and to
have the phase $2\pi k$ are governed by
\begin{eqnarray}\label{generaleqpk1}
\dot
p_{1,k}&=&\JJ_t^{2\to 1}[\pv_{k-1}]-\JJ_t^{1\to 2}[\pv_{k}]\\
\dot p_{2,k}&=&\JJ_t^{1 \to 2}[\pv_{k}]-\JJ_t^{2 \to 1}[\pv_{k}]
\label{generaleqpk2}
\end{eqnarray}
These equations are similar to eqs. (\ref{generaleqp1}) and
(\ref{generaleqp2}), however the probability influx into state 1 for a given phase
$2\pi k$ comes now from states with the phase $2\pi (k-1)$.

The mean phase as well as the mean square phase are given in terms of
the probabilities $\pv_{k}$ by
\begin{eqnarray*}
  \mean{\phi(t)}&=&\sum_{k=-\infty}^\infty 2\pi k ( p_{1,k}(t)
  +p_{2,k}(t))\\
  \mean{\phi^2(t)}&=&\sum_{k=-\infty}^\infty 4\pi^2 k^2 ( p_{1,k}(t)
  +p_{2,k}(t))
\end{eqnarray*}
The instantaneous mean phase velocity $\omega(t)$ 
and instantaneous mean phase diffusion $D_\text{eff}(t)$ are then defined as
\begin{eqnarray}
  \omega(t)&=&\frac{d}{dt}\mean{\phi(t)}\label{defomegat}\\
  D_\text{eff}(t)&=&\frac{1}{2}\frac{d}{dt}\Big[\mean{\phi^2(t)}-\mean{\phi(t)}^2\Big]\label{defD0t}. 
\end{eqnarray}
Asymptotically, i.e. for the initial time $t_0\to -\infty$, the phase
$\phi=2\pi k$ will undergo a diffusional motion \cite{talkner2} with
periodically varying effective phase velocity $\omega(t)$ and
effective diffusion coefficient $D_\text{eff}(t)$.  In this asymptotic regime the
mean phase velocity eq. (\ref{definitionomega0}) and effective phase
diffusion constant eq. (\ref{definitionD0}) can be expressed as the
time average over one period of the external driving of the time
dependent phase velocity and diffusion constant,
\begin{eqnarray}\label{averaged}
\bar \omega=\frac{1}{T}\int_0^T dt \omega(t)\quad\text{and} \quad 
\bar D_\text{eff}=\frac{1}{T}\int_0^T dt D_\text{eff}(t)
\end{eqnarray}
Although the phase velocity and effective phase diffusion constant
eqs. (\ref{defomegat}) and (\ref{defD0t}) have a periodic asymptotic
behavior, the equations (\ref{generaleqpk1}) and (\ref{generaleqpk2}) 
which govern the probabilities $p_{1,k}$ and $p_{2,k}$ obviously 
have no asymptotic solutions.

\section{General theory}
\label{sec.3}
Our aim is to relate the asymptotic phase velocity and effective phase
diffusion constant eqs. (\ref{averaged}) to the microscopic dynamics
eqs. (\ref{generaleqpk1}) and (\ref{generaleqpk2}).  To this end we
introduce a continuous phase distribution $\PP(\phi,t)$ as the
envelope of the discrete phase distribution $p_{1,k}$ and $p_{2,k}$
\cite{freund2,harms} by defining its values at integer multiples of
$2\pi$ as
\begin{eqnarray}\label{phasedistribution}
\PP(\phi=2\pi k,t):=p_{1,k}(t)+p_{2,k}(t).
\end{eqnarray}
The diffusional motion of the phase $\phi$ requires its distribution
$\PP(\phi)$ to  obey the Fokker-Planck equation
\begin{eqnarray}\label{gaussian}
  \pde{t}\PP(\phi,t)=\pde{\phi}(-\omega(t)+
D_\text{eff}(t) \pde{\phi})\PP(\phi,t).
\end{eqnarray}
To establish the relation  between $\omega(t)$ and $D_\text{eff}(t)$ and 
the microscopic dynamics eqs.(\ref{generaleqpk1}) and
(\ref{generaleqpk2}) we 
expand $p_{1,k}$ and $p_{2,k}$  according to
\begin{eqnarray}
\label{ansatz}
  p_{i,k}(t)&=&\sum_{n=0}^\infty q_i^{(n)}(t)
  \pd{n}{\phi}\PP(\phi,t)\Big|_{\phi =2\pi  k},\quad  i=1,2\quad
\end{eqnarray}
This expansion describes how the probability to be in state 1 or 2 for a given
phase $k$ at time $t$, $p_{1,k}(t)$ and $p_{2,k}(t)$ respectively,
is related to the total probability to have a phase $2\pi k$,
$\PP(2\pi k,t)$ and its gradients.

The total probability $p_{1,k}(t)+p_{2,k}(t)$ to have a phase $2\pi k$
neglecting the internal state 1 or 2 is related to the continuous
phase distribution by the defining eq. (\ref{phasedistribution}),
which in turn implies
\begin{eqnarray}
q_1^{(0)}(t)+q_2^{(0)}(t)&=&1\label{normcond}\\
q_1^{(n)}(t)+q_2^{(n)}(t)&=&0 \quad \text{for } n\ge 1\nonumber.
\end{eqnarray}
Inserting the Ansatz eq. (\ref{ansatz}) into the
master eq. (\ref{generaleqpk1}) and (\ref{generaleqpk2}), using
the Fokker-Planck equation (\ref{gaussian}) for the phase and
considering the coefficients of 
the different derivatives $\partial^n/\partial \phi^n \PP(\phi,t)$
eventually leads to (cf. appendix \ref{herleitung})
\begin{widetext}
\begin{eqnarray}
\dot\qv^{(0)}&=&\LT_t[\qv^{(0)}]\label{eps0}\\
\dot \qv^{(1)}&=&\LT_t[\qv^{(1)}+c^{(1)}_t \qv^{(0)}]
-2\pi\JJJ^\text{in}_{t}[\qv^{(0)}]+\omega(\cdot)\qv^{(0)}
\label{eps1}\\
\dot \qv^{(2)}&=&\LT_t[
\qv^{(2)}+c^{(1)}_t \qv^{(1)}+c^{(2)}_t\qv^{(0)}
]-2\pi\JJJ^\text{in}_{t}[\qv^{(1)}+(c_t^{(1)}-\pi)\qv^{(0)}]
+\omega(\cdot)\qv^{(1)}
-D_\text{eff}(\cdot)\qv^{(0)}\label{eps2}
\end{eqnarray}
\end{widetext}
where we have introduced the master operator
\begin{eqnarray*}
\LT_{t}[\cdot]=
\begin{pmatrix}
  \JJ_t^{2\to1}[\cdot]-\JJ_t^{1\to2}[\cdot]\\
  \JJ_t^{1\to2}[\cdot]-\JJ_t^{2\to1}[\cdot]
\end{pmatrix}.
\end{eqnarray*}
The  operator 
\begin{eqnarray*}
\JJJ^\text{in}_{t}[\cdot]=
\begin{pmatrix}
  \JJ_t^{2\to1}[\cdot]\\
  0
\end{pmatrix}
\end{eqnarray*}
accounts for the influx into state 1 and we introduced
\begin{eqnarray*}
c^{(1)}_t(t')&=&\int_{t'}^t d\tau \omega_0(\tau)\\
c^{(2)}_t(t')&=&-\int_{t'}^t
d\tau D_0(\tau)+\omega_0(t)\int_{t'}^t d\tau (t-\tau) \omega_0(\tau).
\end{eqnarray*}

$\qv^{(0)}$ in eq. (\ref{eps0}) shows the same dynamics as $\pv$ in
the two state system without phase eqs. (\ref{generaleqp1}) and (\ref{generaleqp2}), which one
would also expect as this term corresponds to
 an equipartition of phases $\PP(\phi,t)=\text{const}$ in the
 expansion eq. (\ref{ansatz}).
The higher order terms $\qv^{(n)}$ are corrections which emerge due to the fact
that we are considering a non equipartition of phases resulting in
drift and diffusion.

Interestingly if the action of the flux operators on the probabilities
is local in time, i.e. in the Markovian case the terms containing the
$c_t^{(i)}$ are zero, as $c_t^{(i)}(t)=0$, and therefore the dynamics of the
$\qv^{(i)}$ considerably simplifies.

By summing up both components of the vectorial eqs. (\ref{eps1}) and
(\ref{eps2}), using  the
normalization condition eq. (\ref{normcond}) and the fact that 
$(\LT_t)_1+(\LT_t)_2=0$
we arrive at
\begin{eqnarray}
\omega(t)&=&2 \pi \JJ^{2\to1}_{t}[\qv^{(0)}](t)\label{generaleqomega0}\\
D_\text{eff}(t)&=& 
2 \pi\JJ^{2\to1}_{t}[-\qv^{(1)}+(\pi- c_t^{(1)})
\qv^{(0)}](t)\nonumber\\
&=&\pi \omega(t)+2 \pi\JJ^{2\to1}_{t}[-\qv^{(1)}- c_t^{(1)}\qv^{(0)}](t)
\label{generaleqD0}
\end{eqnarray}
The asymptotic mean phase velocity $\bar \omega$ and the asymptotic
effective phase diffusion constant $\bar D_\text{eff}$ can then be determined
from the asymptotic (cyclo stationary) solutions of eqs. (\ref{eps0})
and (\ref{eps1}).  Therefore, the calculation of the asymptotic
effective diffusion constant is reduced to the solution of a cyclo
stationary problem, which in general is simpler that solving the whole
non stationary equations (\ref{generaleqpk1}) and (\ref{generaleqpk2})
with some initial conditions and then taking the asymptotic limit in
eq. (\ref{definitionD0}).

In the following the mean phase velocity and effective phase diffusion
constant will be considered for two different models, namely a
Markovian model \cite{wiesenfeld}, which approximates bistable systems
and a non-Markovian model \cite{prager2}, which serves as an
approximate description for excitable systems.  For the dichotomically
driven Markovian case the mean phase velocity and effective phase
diffusion constant can be explicitly calculated, while for the
non-Markovian case solutions can only be obtained numerically.

\section{A Markovian two state model}\label{bistable}
We  consider now a Markovian two state system with periodically
modulated rates $\gamma_2(t)$
and  $\gamma_1(t)$. Its flux operator $\JJ^{1\to2}$  and 
$\JJ^{2\to1}$ are given by
\begin{eqnarray*}\label{eqp}
  \JJ_t^{1\to2}[\pv](t)=\gamma_1(t) p_1(t)
\quad \text{and} \quad 
  \JJ_t^{2\to1}[\pv](t)=\gamma_2(t) p_2(t)
\end{eqnarray*}
In this Markovian case, the equations, which govern the evolution of
$\qv^{(i)}(t)$
greatly simplify due to the fact that $c^{(i)}_t(t)=0$.
Eqs. (\ref{generaleqomega0}) and (\ref{generaleqD0}) reduce to 
\begin{eqnarray*}
\omega(t)&=&2\pi \gamma_2(t)q_2^{(0)}(t)\\
D_\text{eff}(t)&=&2 \pi^2 \gamma_2(t)q_2^{(0)}(t)-2\pi \gamma_2(t) q_2^{(1)}(t).
\end{eqnarray*}
The equations for $q_2^{(0)}(t)$ and $q_2^{(1)}(t)$ are given by
\begin{eqnarray}
  \dot q_2^{(0)}(t)&=&\gamma_1(t) q_1^{(0)}(t)-\gamma_2(t) q_2^{(0)}(t)\label{eqp0m}\\
  \dot q_2^{(1)}(t)&=&\gamma_1(t) q_1^{(1)}(t)-\gamma_2(t)
  q_2^{(1)}(t)
  +\omega(t)q_2^{(0)}(t)\;\;\;\;\label{eqp1m} 
\end{eqnarray}
Eqs. (\ref{eqp0m}) and (\ref{eqp1m}) can be readily
solved by the method of variation
of constants, 
using  $q_1^{(0)}(t)=1-q_2^{(0)}(t)$ and $q_1^{(1)}(t)=-q_2^{(1)}(t)$
(cf. eq. (\ref{normcond})).
The asymptotic periodic solutions eventually read
\begin{eqnarray}
 q_2^{(0)}(t)&=&
\frac{\int_0^T d\tau  \gamma_1(t-\tau)\exp(-s(\tau,t))}{1-\exp(-s(T,t))}
\label{q20}\\
q_2^{(1)}(t)&=&\frac{\int_0^T d\tau  \omega(t-\tau)
  q_2^{(0)}(t-\tau)\exp(-s(\tau,t))
}{1-\exp(-s(T,t))}\;\;\;\label{q21}
\end{eqnarray}
where $s(\tau,t):=\int_{t-\tau}^t d\tau'
(\gamma_1(\tau')+\gamma_2(\tau'))$. Note that $s(T,t)$ does no longer
depend on $t$.

For a dichotomic symmetric driving with period $T= 2\pi/\omega$,
\begin{eqnarray*}
  \gamma_1(t)=\left\{ 
    \begin{array}{ll}
      r_1&\text{if } t \in [ n T,
      (n+\frac{1}{2})T)\\
      r_2&\text{if } t \in [ (n+\frac{1}{2})T,
      (n+1)T)
    \end{array}
\right.
\end{eqnarray*}
and vice versa for $\gamma_2(t)$ eqs. (\ref{q20}) and (\ref{q21}) can be readily
evaluated leading after some cumbersome algebra to the mean phase
velocity and effective phase diffusion constant 
\begin{eqnarray}\label{finalomega0}
 \bar \omega
&=&\omega_0 +\alpha \Omega \tanh R
\end{eqnarray}
and 
\begin{eqnarray}\label{D0final}
  \bar D_\text{eff}&=&\pi \omega_0 \big[\frac{1}{2}+\alpha(\frac{1}{2}+\cosh^{-2}R)\big]
+\\&&
\pi \alpha \Omega\tanh R\big[-1+\alpha(\frac{1}{2}\cosh^{-2}R+1)\big]\nonumber
\end{eqnarray}
where we have introduced the mean phase velocity without driving
$\omega_0:=2\pi/(\frac{1}{r_1}+\frac{1}{r_2})$, a quantifier for the
driving strength $\alpha=\frac{(r_1-r_2)^2}{(r_1+r_2)^2}$ and some
ratio between inner time scale and driving frequency
$R=\frac{\pi(r_1+r_2)}{2\Omega}$.


Without signal, i.e. $\alpha=0$ eq. (\ref{D0final}) reduces to $\bar D_\text{eff}=\pi\omega_0$,
which agrees with the result in \cite{cox}, 
$\bar D_\text{eff}=(2\pi)^2/2(\mean{t^2}-\mean{t}^2)/\mean{t}^3$.

Next we consider the small and large noise limits of the phase
velocity $\bar \omega$ and phase diffusion constant $\bar D_\text{eff}$  for the case
of Arrhenius  rates $r_{1/2}=r_0
\exp(-\frac{\Delta U\pm A}{D})$.
In this case $\alpha=\tanh^2(\frac{A}{D})$.

If for a fixed driving frequency the noise level is sufficiently small
such that $R\ll 1$
eqs. (\ref{finalomega0}) and (\ref{D0final}) reduce to
\begin{eqnarray*}
  \bar \omega&\approx& \omega_0+\alpha\Omega R=
  \frac{\pi}{2}(r_1+r_2)\approx\frac{\pi}{2}r_2 \\
\bar D_\text{eff}&\approx& \pi \omega_0(\frac {1}{2}+\frac{3}{2}\alpha)
+\pi\alpha\Omega R(-1+\frac{3}{2}\alpha)
\\&=&\frac{\pi^2}{4}(r_1+r_2)\approx
\frac{\pi^2}{4} r_2
\end{eqnarray*}
where in the last step we used the fact that $r_2$ dominates $r_1$
for small noise levels.
Therefore, at the level of
phase velocity and phase diffusion, the process  behaves like
a process without driving whose rates are both equal to $\frac{r_2}{2}$.

On the other hand if the noise level is large  
and the driving frequency is small compared to $\alpha_0$ such that
$R\gg 1$ we get
\begin{eqnarray*}
  \bar \omega&\approx&\omega_0+\alpha \Omega= 2\pi\frac{r_1r_2}{r_1+r_2}+
\Omega\frac{(r_1-r_2)^2}{(r_1+r_2)^2}\\
\bar D_\text{eff}&\approx& \frac{\pi}{2}\omega_0(1+\alpha)+\pi\alpha\Omega(-1+\alpha)
\\&=&
2\pi^2\frac{r_1r_2(r_1^2+r_2^2)}{(r_1+r_2)^3}-4 \pi \Omega\frac{(r_1-r_2)^2}{(r_1+r_2)^4}
\end{eqnarray*}
The first terms in these  expressions correspond to a process without driving with one
rate equal to $r_1$ and the other equal to $r_2$, while the second
terms are
corrections which vanish for vanishing driving frequency.

Between these regions we have a competing behavior. If for a fixed
driving amplitude $A$, the noise strength $D$ is sufficiently small, such
that $\alpha \approx 1$ and $\omega_0\approx 0$, and simultaneously, for a fixed driving frequency $\Omega$,
$D$ is sufficiently large such that
$R\gg 1$, i.e. $\tanh R\approx 1$  we have
\begin{eqnarray*}
  \bar\omega&\approx& \Omega \\
\bar D_\text{eff}&\approx& 0
\end{eqnarray*}
i.e. frequency and phase locking occur.

Having calculated the effective diffusion coefficient and the mean
phase velocity we can evaluate the P\'eclet number
\begin{eqnarray}\label{pecletdef}
{\rm Pe}:=\frac{2\pi\bar \omega}{\bar D_\text{eff}}\label{eq:2}
\end{eqnarray}
which is a measure of the phase coherence.

\begin{figure}[htb p]
\centerline{\includegraphics[width=9cm]{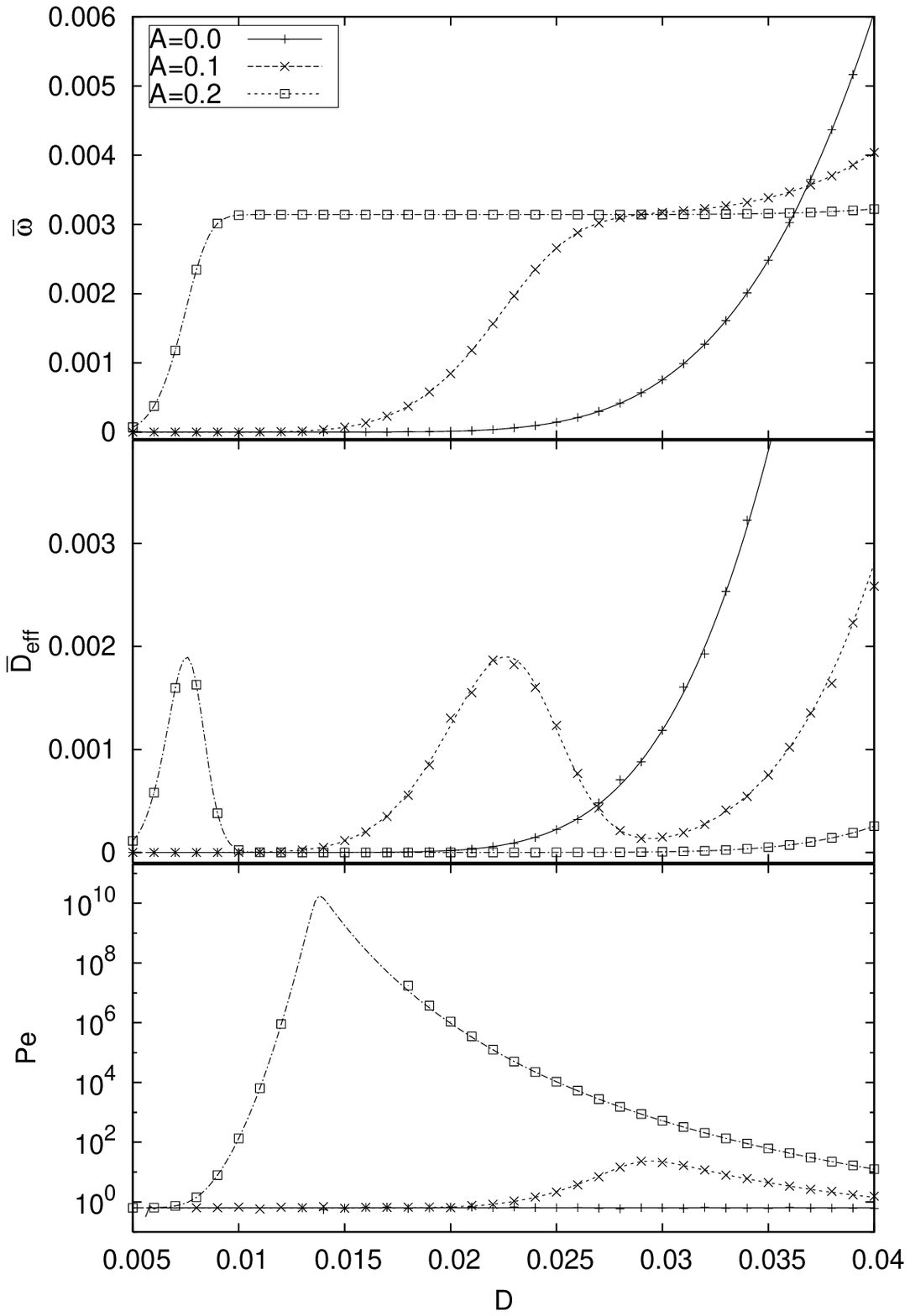}}
\caption{\label{picomega0}
  Mean phase velocity $\omega_0$ (top), effective phase diffusion
  constant $D_0$ (middle) and P\'eclet number ${\rm Pe}$ (bottom) of
  the Markovian model for
  different values of the driving amplitude.  \\Symbols are simulation
  data of the two state system, lines according to eq.
  (\ref{finalomega0}), (\ref{D0final}) and (\ref{pecletdef}), respectively.  Other
  parameters: $r_0=1$ and $\Delta U=0.25$, $\Omega=0.001\pi$.  The
  deviation between theory and simulations in the P\'eclet number for
  low noise intensities is due to limited simulation time.  }
\end{figure}

In Fig.\ref{picomega0} the theoretical results
eqs. (\ref{finalomega0}),(\ref{D0final}) and
(\ref{pecletdef}) are compared to
simulations of the driven two state system. To compute these results we have modified an
algorithm presented in \cite{gibson} taking into account that the transition
rates are piecewise constant in time due to the dichotomic driving.
Let us assume we start at time $t$ in state $1$ and the input defines
the rate to have the value $r_1$.  Then we draw a random number $\tau$
according to the corresponding waiting time distribution $w_{r_1}(\tau)=r_1\exp(-r_1
\tau)$.  If $t+\tau$ is smaller then the time $t_s$ of the next switching of
the input we set the running time to $t+\tau$ and perform the
transition to the second state of the system. This state $2$ will be
left with rate $r_2$ and we proceed accordingly. Contrary if during
the interval $[t,t+\tau]$ a switching of the input occurs we set the
running time equal to the switching time $t_s$ but remain in state $1$. After
switching of the input the rate for leaving state $1$ is now $r_2$ and
we proceed by drawing a new waiting time according to the new density
$w_{r_2}(\tau)=r_2\exp(-r_2 \tau)$.

\begin{figure}[htbp]
\centerline{\includegraphics[width=9cm]{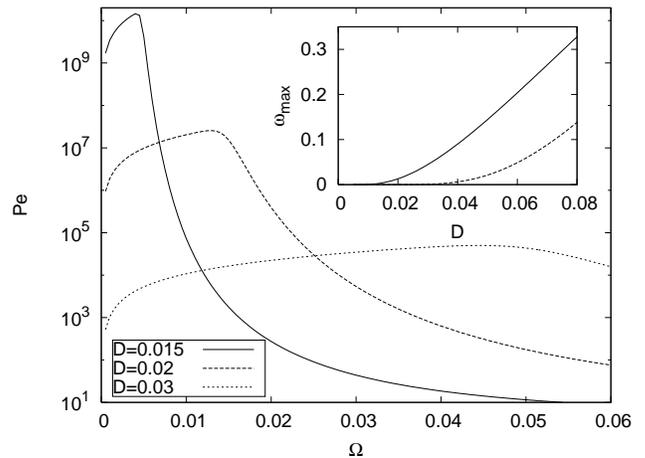}}
\caption{\label{picbfresonance}
  P\'eclet number ${\rm Pe}$ of the Markovian model as a function of driving frequency
  $\Omega$ for different noise values. $A=0.2$, other parameters as in
  Fig.  \ref{picomega0}.  The inset shows the driving frequency at the
  maximum P\'eclet number as a function of noise strength (solid line)
  compared to the intrinsic frequency $\Gamma$ without driving ($A=0$)
  (dashed line).
}
\end{figure}

The P\'eclet number shows a maximum as a function of noise strength,
indicating stochastic resonance. For a strong driving, it varies over
several orders of magnitude with varying noise strength $D$.
Interestingly the P\'eclet number shows also a non monotonic behavior as a
function of driving frequency for a fixed noise level, i.e. using this
number as a measure of the quality of the response to the external
signal we discover a ``bona fide `` resonance
(Fig. \ref{picbfresonance}).

\section{Excitable systems} \label{excitable}
In this section we consider the phase velocity and diffusion of a non
Markovian model \cite{prager2}.
This two state model mimics the dynamics of an excitable systems by
dividing it into an excitation step  and the evolution along the excitation loop.
Its dynamics is given by
\begin{eqnarray}\label{eqp1excitable}
  \dot p_1(t)&=&\gamma(t) p_2(t)-\int_{t_0}^{t} d\tau w(t-\tau)\gamma(\tau) p_2(\tau)\\
\label{eqp2excitable} 
  \dot p_2(t)&=&-\gamma(t) p_2(t)+\int_{t_0}^{t} d\tau w(t-\tau)\gamma(\tau) p_2(\tau).\;\;
\end{eqnarray}
with initial conditions 
\begin{eqnarray}\label{initialcond}
p_1(t_0)=0 \quad \text{ and } \quad p_2(t_0)=1.
\end{eqnarray}
State 2 represents the rest state, in which we start at initial time $t_0$
>From there  the system is excited
due to noise and the external periodic subthreshold signal, leading 
a rate process with  rate $\gamma(t)$, which depends
periodically on time. 
This Markovian excitation step is described by 
\begin{eqnarray}\label{fluxopexcitable1}
  \JJ_t^{2\to1}[\pv](t)&=&\gamma(t) p_2(t).
\end{eqnarray}
State 1 accounts for the motion on the
excitation loop on which the systems spends a time distributed according to the
waiting time distribution $w(\tau)$, which is assumed not to depend
on the weak external driving. The flux
from state 1 back to state 2 is then expressed in terms of the
flux from state 2 to state 1 at prior times $\tau$ 
between $t_0$ to $t$  $\gamma(\tau) p_2(\tau)$, which renders the
description non Markovian, leading to the flux operator
\begin{eqnarray}\label{fluxopexcitable2}
  \JJ_t^{1\to2}[\pv](t)&=&\int_{t_0}^t d\tau  w(t-\tau) \gamma(\tau) p_2(\tau).
\end{eqnarray}
Note that this operator depends explicitly on the initial time $t_0$.

To calculate the asymptotic periodic solution it will
be useful to first formally integrate  eqs. (\ref{eqp1excitable}), (\ref{eqp2excitable}) 
taking into account the initial conditions (\ref{initialcond}) and then
taking the initial time $t_0$ to $-\infty$. The resulting equations are
\begin{eqnarray}\label{eqp1excitableasymp}
  p_1(t)&=&\int_{0}^{\infty} d\tau z(\tau)\gamma(t-\tau) p_2(t-\tau)\\
\label{eqp2excitableasymp}  
  p_2(t)&=&1-\int_{0}^{\infty} d\tau z(\tau)\gamma(t-\tau) p_2(t-\tau).
\end{eqnarray}
where $z(\tau)=1-\int_0^\tau d\tau' w(\tau')$ is the probability to
spent a time longer than $\tau$ on the excitation loop.  By
differentiating these equations with respect to $t$ one recovers the
original eqs. (\ref{eqp1excitable}) and (\ref{eqp2excitable}) in the
limit $t_0\to -\infty$ \cite{comment}.

If we take into account the phase eq. (\ref{eqp2excitableasymp})
has  to be replaced by
\begin{eqnarray}\label{p2vonp1k}
  p_{(1,k)}(t)&=&\int_0^\infty d\tau z(\tau) \gamma(t-\tau) p_{(2,k-1)}(t-\tau).
\end{eqnarray}
We also have to take care of the flux operator $\JJ_t^{1\to2}$ which
in the asymptotic case is given by (cf. eq. (\ref{fluxopexcitable2}))
\begin{eqnarray}\label{fluxopexcitable2asymp}
  \JJ_t^{1\to2}[\pv](t)&=&\int_{0}^\infty d\tau  w(\tau) \gamma(t-\tau) p_2(t-\tau).
\end{eqnarray}

\begin{figure}[htbp]
  \centerline{\includegraphics[width=9cm]{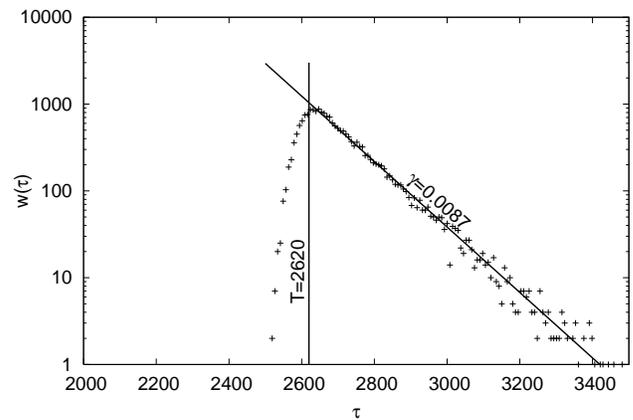}}
\caption{\label{wtds}
  Inter spike interval distribution of the FHN system eqs.
  (\ref{fhneqs}) with constant signal $s(t)=0.04$ for a low noise
  level $D=10^{-5}$ and strong time scale separation $\epsilon=0.001$.
  Other parameters see text.  }
\end{figure}

In the following we assume a fixed waiting time $T$ on the excitation
loop, i.e. $w(\tau)=\delta(T-\tau)$ and $z(\tau)=\theta(T-\tau)$.
Such an assumption is justified in the low noise limit for e.g.
FitzHugh-Nagumo models (cf. Fig. \ref{wtds}). In this case eq.
(\ref{fluxopexcitable2asymp}) simplifies to
\begin{eqnarray}\label{fluxopexcitable2asympsimple}
  \JJ_t^{1\to2}[\pv](t)&=&\gamma(t-T) p_2(t-T).
\end{eqnarray}

Then, according to eqs. (\ref{generaleqomega0}) and
(\ref{generaleqD0}), the time dependent phase velocity $\omega_0(t)$
and effective phase diffusion constant $D_0(t)$ are given by
\begin{eqnarray}
  \omega(t)&=& 2\pi \gamma(t)q_2^{(0)}(t)\label{omega0excitable}\\
  D_\text{eff}(t)&=&
- 2\pi \gamma(t)q_2^{(1)}(t)\label{D0excitable}+2\pi^2   \gamma(t)q_2^{(0)}(t), \nonumber
\end{eqnarray}
which are the same expressions as in the Markovian case, as the flux
operator $\JJ_t^{2\to1}$ is the same. However the equations governing
the $\qv^{(i)}$ are different.  Following the same procedure we used
to treat eqs.  (\ref{generaleqpk1}) and (\ref{generaleqpk2}) eq.
(\ref{p2vonp1k}) together with normalization the condition
(\ref{normcond}) leads to
\begin{eqnarray}
  1-q_2^{(0)}(t)&=&\int_0^T d\tau 
  \gamma(t-\tau)q_2^{(0)}(t-\tau)
\label{normcondp20}\\
  -q_2^{(1)}(t)&=&\int_0^T d\tau \gamma(t-\tau)q_2^{(1)}(t-\tau)\label{normcondp21}\\
&&\hspace{-2cm}+\int_0^T d\tau   \gamma(t-\tau)q_2^{(0)}(t-\tau)(\int_0^\tau d\tau' \omega(t-\tau')-2\pi)\nonumber.
\end{eqnarray}
The periodic solutions of eqs. (\ref{normcondp20}),
(\ref{normcondp21}) can be numerically obtained in Fourier space using
a linear solver like LAPACK.

To investigate the role of noise on the synchronization in excitable
system we choose an Arrhenius type excitation rate for the transition 
from the rest state
$2$ onto the excitation loop $1$. We further assume that the external
driving acts as a modulation of the potential barrier. 
Again we consider  a dichotomic periodic
driving, i.e. the excitation rate $\gamma(t)$ periodically switches between the two values
$r_1=r_0 \exp(-(\Delta U-A)/D)$ and $r_2=r_0 \exp(-(\Delta U+A)/D)$.

\begin{figure}[htbp]
\centerline{\includegraphics[width=9cm]{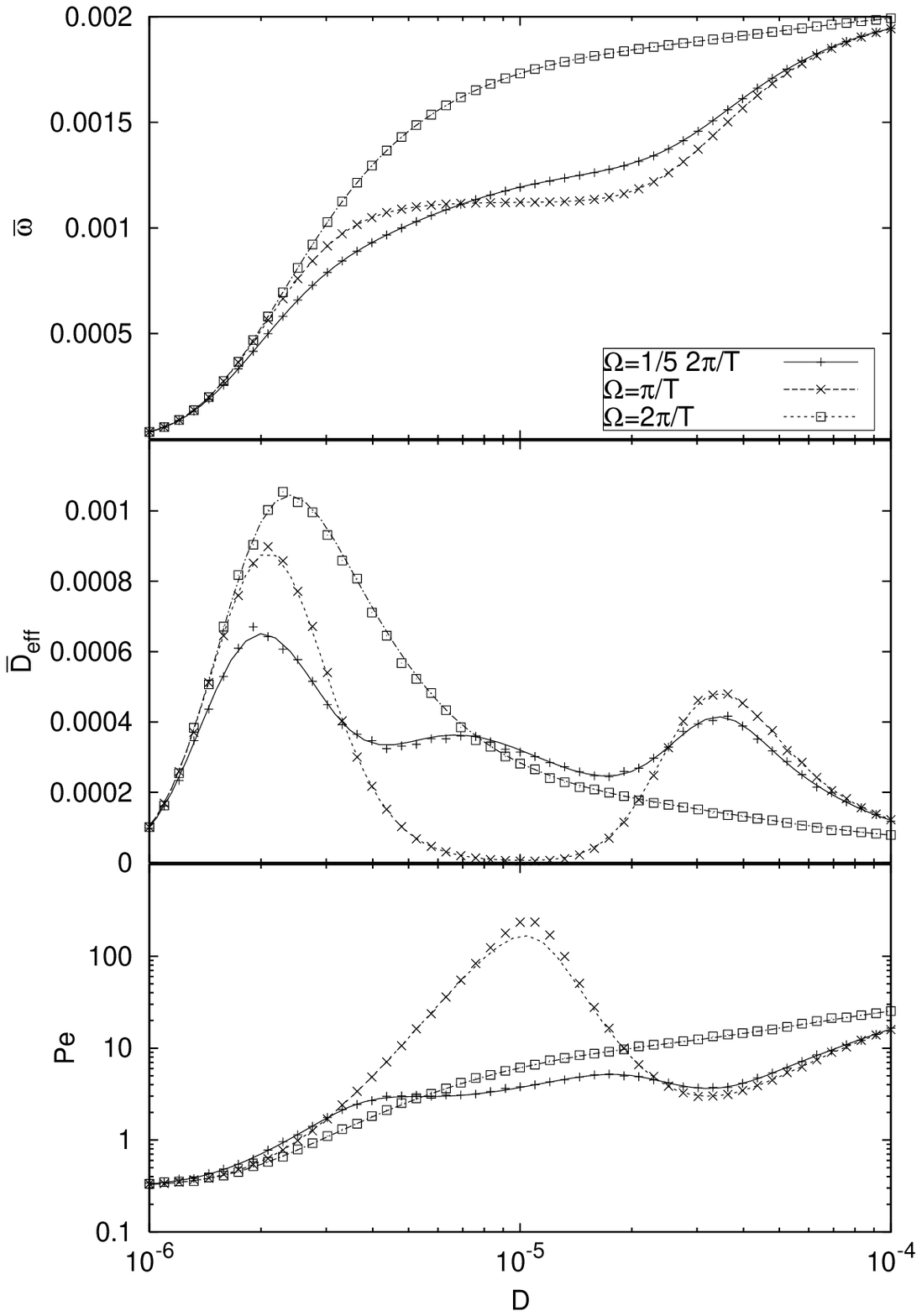}}
\caption{\label{comptheotwostateexcitable}
  Mean phase velocity $\bar \omega$ (top, inset), effective phase
  diffusion constant $\bar D_\text{eff}$ (top) and P\'eclet number ${\rm Pe}$
  (bottom) of the non Markovian model for different values of the driving frequency $\Omega$.
  \\Symbols are simulation data of the two state system, lines
  according to numerical evaluation of the theory.  Other parameters:
  $T=2800$, $r_0=0.0044$, $\Delta U=5.6\cdot 10^{-5}$ and $A=5.0 \cdot 10^{-5}$.  }
\end{figure}

The resulting phase velocity, effective phase diffusion and P\'eclet
number as a function of noise strength $D$ are shown in Fig.
\ref{comptheotwostateexcitable}.  As in the case of bistable systems
we observe frequency and phase locking, however there exist preferred
driving frequencies for which high synchronization is achieved and
other frequencies which show no synchronization at all.

The P\'eclet number shows a local maximum at a
finite noise strength. Contrary to the bistable situation however, 
the phase diffusion constant decreases again and the P\'eclet
number therefore increases for large noise levels.
This behavior is originated in  the fixed time $T$ on the excitation loop.
Taking into account the high rate and therefore small
waiting time and variance of the excitation step for
high noise levels this leads to a low variance of the spiking, which
implies a low diffusion of the phase.
We mention that this low phase diffusion does not imply
synchronization since the frequencies are not locked. Also we note that
in real excitable systems the behavior differs. For higher noise
levels the time spent on the excitation loop will have a variance in
these systems which yields an increasing phase diffusion with growing noise.

\begin{figure}[htbp]
\centerline{\includegraphics[width=9cm]{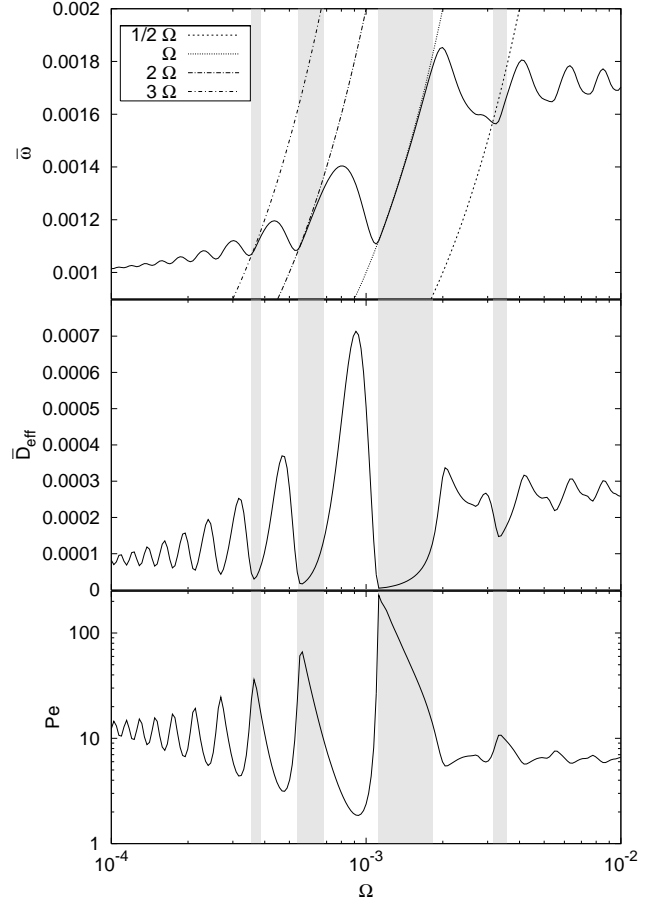}}
\caption{\label{picomegadependence}
  Mean phase velocity $\bar \omega$ (top), effective phase diffusion
  constant $\bar D_\text{eff}$ (middle) and P\'eclet number
  $\text{Pe}$ (bottom) of the non Markovian model as a
  function of driving frequency $\Omega$ for $D=0.00001$.  The shaded
  regions are a guide for the eye and represent regions of frequency
  synchronization.  In these regions we also find a small effective
  phase diffusion and therefore a high P\'eclet number.  Other
  parameters: $T=2800$, $r_0=0.0044$, $\Delta U=5.6\cdot  10^{-5}$ and $A=5.0 \cdot 10^{-5}$.}
\end{figure}

\begin{figure}[htbp]
\centerline{\includegraphics[width=10.5cm]{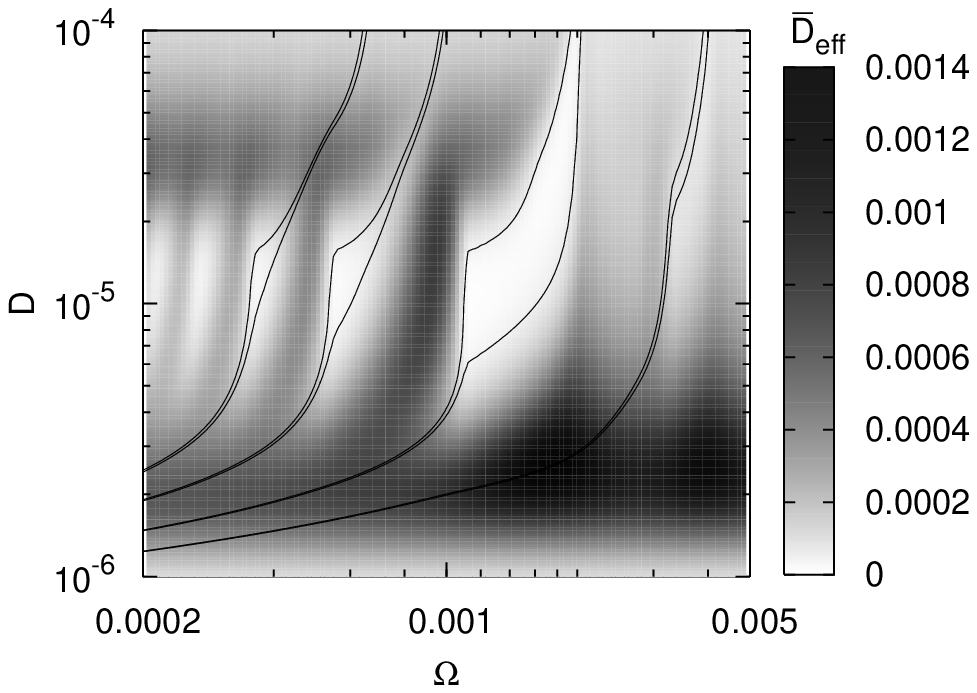}}
\caption{\label{picsynchronizationregions}
Effective phase diffusion constant ${\bar D_\text{eff}}$ of the non
Markovian model,
as a function of driving frequency $\Omega$ 
and noise level $D$. 
The black lines show regions of frequency locking 
$(1-\epsilon)\Omega<n \bar \omega<(1+\epsilon)\Omega$, $\epsilon=0.01$
with (from left to right) $n=3,2,1,\frac{1}{2}$.
These regions of frequency locking coincide with low phase diffusion.
Other parameters as in Figs. \ref{comptheotwostateexcitable} and 
\ref{picomegadependence}. 
}
\end{figure}

As seen in Fig.\ref{comptheotwostateexcitable} the synchronization
behavior strongly depends on the driving frequency. To further
analyze this effect we have plotted in Fig.\ref{picomegadependence}
the mean phase velocity, phase diffusion coefficient and the P\'eclet
number as function of the driving frequency. They show a complex sequence of
different locking regions between the driving and the system's
response \cite{parmananda,lee}, represented by
shaded regions.
In these locking regions the effective phase diffusion is small
(see Fig.\ref{picsynchronizationregions}).
We mention that the maximal frequency of the excitable system is
$\bar{\omega}_{\rm max} = 2\pi / T$ where $T$ is the time on the
excitation loop. There can not be $1:1$ synchronization for $\Omega >
\bar{\omega}_{max}$.

Let us for a moment assume the extreme case where one excitation rate
$r_1$ is infinity and
the other $r_2$ is zero. Then the system remains in the rest state as long
as the input causes the vanishing excitation rate.  After the input
changes the system immediately starts with the excitation loop where
it stays the time $T$.  For a $1:1$ locking this time $T$ must be
larger than half the period but smaller the full period $2 \pi/\Omega$
of the driving.  Otherwise, if the duration of the excitation loop
would be smaller than half the period the system returns to the rest
state where it immediately starts a new excitation. In consequence
$1:n$ locking where the output frequency is $n$ times higher than the
input frequency occurs if the period of the driving is between
$(n-1/2)T$ and $nT$.

The opposite case  where a fast input locks a slow output
occur if multiple periods of the input fit into the excitation time.
During the excitation the system does not respond to the changes of the
input.  If  the input has the phase with long waiting
time after the system has completed the excitation loop, it has to
wait until the input changes to the phase with the small waiting time,
leading to a $n:1$ synchronization where $n$ is the number of signal
periods which fit into the excitation time $T$.

However if the system finds the high excitation rate after excursion it
 immediately starts a new excitation loop and repeats these until it will find the
phase with long waiting times. This yields a $n:m$ frequency locking
with $n>m$. Note that there are no $n:m$ locking modes with $n<m$
except the $1:m$ modes described above.

Realistic noise dependent time scales will weaken the
extreme behavior of the  situation considered above. 
There are two competing effects namely increasing the noise increases $r_1$ as well
as $r_2$ while decreasing the noise increases the ratio between $r_1$
and $r_2$ and therefore the effect of the driving.
Hence, we find synchronization in a finite window of noise intensities where the two
activation times enclose the time $T$  on the excitation loop, 
\begin{eqnarray}
\frac{1}{r_1}\ll T
\ll \frac{1}{r_2}. \label{ratecond}
\end{eqnarray}
We point out that this latter time plays the essential role within the
synchronization process, i.e. this time scale and the period of the
external drive have to be tuned appropriately to get phase
synchronization.  
Noise as well as the amplitude of driving define the two excitation
rates and have to be chosen such that eq. (\ref{ratecond}) is
optimally fulfilled, i.e. that the input acts as much as possible as a on-off switch on the
excitation process. A deviation from this extremal behavior
leads to a narrowing of the driving frequency windows amenable to
frequency locking and a shift of these windows to lower frequencies.


Finally we compare the theory to a dynamical system with
excitable dynamics, namely the FHN model \cite{fhn1,fhn2}
\begin{eqnarray}\label{fhneqs}
\dot x&=&x-x^3-y+\sqrt{2 D} \xi(t)\nonumber\\
\dot y&=&\epsilon(x+a_0-a_1 y-s(t))
\end{eqnarray}
This system is driven by a dichotomic periodic signal $s(t)$ with
values $\pm A$ where $A=0.015$. Setting $a_0=0.405$ and $a_1=0.5$ the
system is in the excitable regime for both values of the signal, i.e.
the signal is a sub-threshold signal. We further consider a strong
time scale separation $\epsilon=0.001$ as well as a small noise level
$D=10^{-5}$.  The phase of the system is defined to increase by
$2\pi$ each time a spike is generated.
From simulations of the inter spike interval
distribution (see Fig. \ref{wtds}) for constant signal $\pm A$ we find
the corresponding parameters of the two state model to be
$T\approx2620$, $r_1\approx 0.0087$ and $r_2\approx 8.3 10^{-8}$ .
  
\begin{figure}
  \centerline{ \includegraphics[width=9cm]{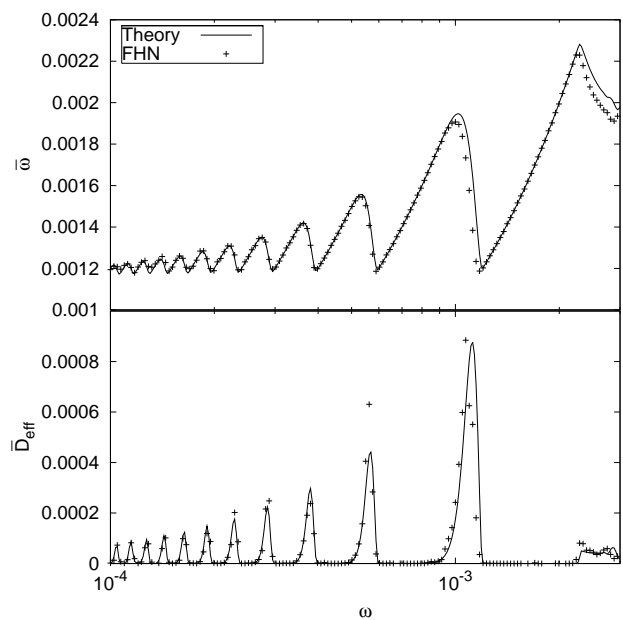}}
\caption{\label{piccomptheoFHN}
  Comparison between theory and FHN system for mean phase velocity
  $\bar \omega$ (top) and effective phase diffusion constant $\bar D_\text{eff}$
  (bottom) as a function of driving frequency.  Other parameters see
  text.  }
\end{figure}

The results for the phase velocity $\bar \omega$ and effective phase
diffusion constant $\bar D_\text{eff}$ for the FHN system (numerical simulation
of eqs. (\ref{fhneqs})) and the theory eqs. (\ref {omega0excitable})
and (\ref {D0excitable}) are shown in Fig. \ref{piccomptheoFHN}.  They
show a good qualitative agreement over a large range of driving
frequencies.  The deviation for larger driving frequencies is due to
the fact that, in contrast to the assumptions of our two state model,
the time $T$ spent on the excitation loop depends if however only
weakly on the driving.

\section{Conclusions}
We have derived a general theory to calculate the asymptotic effective phase
velocity and phase diffusion constant in periodically driven 
two state systems. This theory was applied to two different two state
models, one with Markovian dynamics representing bistable systems 
 and the other with non Markovian dynamics, modeling excitability.

In the Markovian case 
analytical results have been calculated for
dichotomic driving with arbitrary driving amplitudes. We found phase
synchronization for optimal noise intensities if Arrhenius type rates
for the transitions between the states are assumed. The mean frequency
of the system is locked to the frequency of the external stimulus and
the effective phase diffusion coefficient becomes vanishingly small.
The P\'eclet number however shows  a maximum not only as a function of noise
strength but also as a function of  driving frequency, i.e. a  ``bona fide'' resonance. 
Frequency locking occurs as long as the driving frequency is smaller than the maximal
transition rates which are attained for large noise.

In the non Markovian case the phase
velocity, phase diffusion coefficient and P\'eclet number also
prove phase synchronization between input and output.  However the picture differs from the
previous case showing a sequence of frequency locking modes.  These
different regions of locking are accompanied with low phase diffusion.
The main conditions for locking are expressed by relations between the
driving frequency $\Omega$ and the time $T$ spend on the excitation
loop.  The noise dependent and periodically modulated transition rates
from the rest to the excited state act as a switch for the spiking.
$1:n$ locking, i.e. a slow input and fast output, occurs for a certain
window of noise intensities if, in first approximation, the driving
frequency is between $2\pi/\left((n-1/2)T\right)$ and $2\pi/(nT)$,
respectively. For the opposite case of fast input and slow output we
find $1:n$  but also $m:n,\;m<n$ frequency locking. The theoretical
results for the non Markovian model of excitable systems agree well
with simulations of a FitzHugh-Nagumo system.

This work was supported by DFG-Sfb 555. The authors thank P. Talkner
and J. A. Freund for fruitful collaboration.

\appendix
\section{} \label{herleitung}
Our aim is to express the phase distribution
$\PP(\phi-\Delta\phi,t-\tau)$ in terms of $\PP(\phi,t)$ and its
derivatives with respect to $\phi$, $\partial^n/\partial \phi^n
\PP(\phi,t)$.  To this end we start by expanding
$\PP(\phi-\Delta\phi,t-\tau))$ in a Taylor series around $\phi$ and
$t$,
\begin{eqnarray*}
\lefteqn{ \PP(\phi-\Delta\phi,t-\tau)}\\
&=&\sum_{n=0}^\infty\sum_{m=0}^\infty
    \frac{1}{n!m!}\left(\frac{\partial^n}{\partial
        \phi^n}\right)\left(\frac{\partial^m}{\partial t^m}\right) 
\PP(\phi,t)(-\Delta\phi)^n(-\tau)^m
\end{eqnarray*}
To process the time derivatives we use the Fokker-Planck equation eq. (\ref{gaussian})
\begin{eqnarray*}
  \pde{t}\PP(\phi,t)=\pde{\phi}(-\omega(t)+
D_\text{eff}(t) \pde{\phi})\PP(\phi,t),
\end{eqnarray*}
taking care of the explicit time
dependence of $\omega(t)$ and $D_\text{eff}(t)$ which leads to
\begin{widetext}
\begin{eqnarray*}
\lefteqn{ \PP(\phi-\Delta\phi,t-\tau)
= \PP(\phi,t)-
\Big[\Delta \phi+
\sum_{m=1}^\infty
\frac{(-\tau)^m}{m!}\frac{\partial^{m-1}\omega(t)}{\partial
  t^{m-1}}\Big]\pde{\phi}\PP(\phi,t)+}\\&&
 \Big[
\frac{1}{2} \Delta\phi^2
+\sum_{m=1}^\infty \frac{(-\tau)^m}{m!}\frac{\partial^{m-1}D_\text{eff}(t)}{\partial t^{m-1}}
+\omega(t)
\sum_{m=2}^\infty \frac{(-\tau)^m}{m (m-2)!}\frac{\partial^{m-2}\omega(t)}{\partial t^{m-2}}
+\Delta\phi\sum_{m=1}^\infty
\frac{(-\tau)^m}{m!}\frac{\partial^{m-1}\omega(t)}{\partial t^{m-1}}
\Big]\pdz{\phi}\PP(\phi,t)
\\&&+O(3).
\end{eqnarray*}
\end{widetext}
where $O(3)$ denotes third or higher  derivatives of $\PP(\phi,t)$
with respect to $\phi$.
The sums containing the derivatives of $\omega(t)$ and $D_\text{eff}(t)$ can
be further evaluated, leading to
\begin{eqnarray*}
\sum_{m=1}^\infty \frac{(-\tau)^m}{m!}\frac{\partial^{m-1}\omega(t)}{\partial t^{m-1}}
&=&-\sum_{m=0}^\infty
 \frac{1}{m!}\frac{\partial^{m}\omega(t)}{\partial t^{m}}
 \int_0^\tau d\tau'(-\tau')^{m}
\\
&=&-\int_0^\tau d\tau'\omega(t-\tau')
\end{eqnarray*}
 and analogously
 \begin{eqnarray*}
\sum_{m=1}^\infty \frac{(-\tau)^m}{m!}\frac{\partial^{m-1}D_\text{eff}(t)}{\partial t^{m-1}}
&=&-\int_0^\tau d\tau'D_\text{eff}(t-\tau')
 \end{eqnarray*}
and
\begin{eqnarray*}
\sum_{m=2}^\infty \frac{(-\tau)^m}{m (m-2)!}
\frac{\partial^{m-2}\omega(t)}{\partial t^{m-2}}
&=&\int_0^\tau d\tau' \tau' \omega(t-\tau').
\end{eqnarray*}
Thus we eventually arrive at
\begin{eqnarray*}
\lefteqn{\PP(\phi-\Delta\phi,t-\tau)
=}\\
&&\PP(\phi,t)+(c_t^{(1)}(t-\tau)-\Delta\phi)\pde{\phi}\PP(\phi,t)+
\\&&\Big[\frac{1}{2}(\Delta \phi)^2-\Delta\phi
c_t^{(1)}(t-\tau)+c_t^{(2)}(t-\tau)
\Big]\pdz{\phi}\PP(\phi,t)
\\&&+O(3)
\end{eqnarray*}
where 
\begin{eqnarray*}
c_t^{(1)}(t')&=&\int_{t'}^t d\tau \omega(\tau)\\
c_t^{(2)}(t')&=&-\int_{t'}^t
d\tau D_\text{eff}(\tau)+\omega(t)\int_{t'}^t d\tau (t-\tau) \omega(\tau).
\end{eqnarray*}
Next we insert our Ansatz eq. (\ref{ansatz})
\begin{eqnarray*}
  \pv_k(t)&=&\sum_{n=0}^\infty \qv^{(n)}(t)
  \pd{n}{\phi}\PP(\phi,t)\Big|_{\phi =2\pi  k}.
\end{eqnarray*}
into the dynamical equations eqs.(\ref{generaleqpk1}) and (\ref{generaleqpk2})
\begin{eqnarray}\label{dyneqapp1}
\dot
p_{(1,k)}&=&\JJ_t^{2\to 1}[\pv_{k-1}]-\JJ_t^{1\to 2}[\pv_k]\\
\dot p_{(2,k)}&=&\JJ_t^{1 \to 2}[\pv_k]-\JJ_t^{2 \to 1}[\pv_k]\label{dyneqapp2}.
\end{eqnarray}
Using again the Fokker-Planck equation eq. (\ref{gaussian})
the left hand side of eqs. (\ref{dyneqapp1}) and (\ref{dyneqapp2}) is given by
\begin{widetext}
\begin{eqnarray*}
\tde{t}\pv_k(t)=\sum_{n=0}^\infty \big(\tde{t}  \qv^{(n)}(t)\big) \pd{n}{\phi}
  \PP(\phi,t)+\sum_{n=0}^\infty 
  \qv^{(n)}(t)\pd{n+1}{\phi}(-\omega(t)+ D_\text{eff}(t)\pde{\phi})
  \PP(\phi,t).
\end{eqnarray*}
The different terms on the right hand side of eqs. (\ref{dyneqapp1})
and (\ref{dyneqapp2}) read
\begin{eqnarray*}
\JJ_t^{i\to j}[\pv_{k-1}]&=&
\sum_n \JJ_t^{i\to j}[\qv^{(n)}]\big(\pd{n}{\phi}-2\pi \pd{n+1}{\phi}+2
\pi^2 \pd{n+2} {\phi}\big) \PP(\phi,t)
+\JJ_t^{i\to j}[c_t^{(1)} \qv^{(n)}]\big(\pd{n+1}{\phi}-2\pi \pd{n+2}{\phi}\big)\PP(\phi,t)
\\&&+\JJ_t^{i\to j}[c_t^{(2)} \qv^{(n)}]\pd{n+2}{\phi}\PP(\phi,t)+O(n+3)\\
\JJ_t^{i\to j}[\pv_{k}]&=&\sum_n \JJ_t^{i\to j}[\qv^{(n)}]\pd{n}{\phi} \PP(\phi,t)+
\JJ_t^{i\to j}[c_t^{(1)} \qv^{(n)}]\pd{n+1}{\phi}\PP(\phi,t)
+\JJ_t^{i\to j}[c_t^{(2)} \qv^{(n)}]\pd{n+2}{\phi}\PP(\phi,t)+O(n+3)
\end{eqnarray*}
\end{widetext}

Equating now the coefficients of $\PP(\phi,t)$,
$\pde{\phi}\PP(\phi,t)$
and $\pdz{\phi}\PP(\phi,t)$ finally leads to eqs. (\ref{eps0}) to
(\ref{eps2})


\begin{thebibliography}{10}
\bibitem{benzi} R. Benzi, A. Sutera, A. Vulpiani, J. Phys. A 14,
  L453 (1981).
\bibitem{cnicolis}C. Nicolis and G. Nicolis, Tellus 33, 225 (1981). 
\bibitem{gammaitoni} L. Gammaitoni, P. Hänggi, P. Jung, F. Marchesoni,
  Rev. Mod. Phys. 70, 233 (1998).
\bibitem{uspekhi}V. S. Anishchenko, A. B. Neiman, F. Moss and L.
  Schimansky-Geier, Phys. Usp. 42, 7 (1999).
\bibitem{ani_book} V. Anishchenko, A. Neiman, A. Astakhov, T.
  Vadiavasova, and L. Schimansky-Geier, {\em Chaotic and Stochastic
    Processes in Dynamic Systems}, Springer Verlag,
  Berlin-Heidelberg-New York, Springer-Series on Synergetics (2002).
\bibitem{moss}T. Zhou, F. Moss, and P. Jung, Phys. Rev. A 42, 3161
  (1990).
\bibitem{santucci} L. Gammaitoni, F. Marchesoni, and S. Santucci,
  Phys.  Rev. Lett. 74, 1052 (1995).
\bibitem{choi}M. H. Choi, R. F. Fox, and P. Jung Phys. Rev. E 57, 6335
  (1998).
\bibitem{giacomelli}G. Giacomelli, F. Marin, and I. Rabbiosi Phys.
  Rev. Lett. 82, 675 (1999).
\bibitem{talkner} P. Talkner, Physica A 325, 124 (2003).
\bibitem{strato}R. L. Stratonovich, {\em Topics in the theory of
    random noise}, vol 2, Gordon and Breach (1967).
\bibitem{neiman} A. Neiman, A. Silchenko, V. Anishchenko, L.
  Schimansky-Geier, Phys. Rev. E 58, 7118 (1998).
\bibitem{collins} A. Neiman, L. Schimansky-Geier, F. Moss, B. Shulgin,
  and J.J. Collins, Phys. Rev. E 60, 284 (1999).
\bibitem{lindner_rep}B. Lindner, J. Garcia-Ojalvo, A. Neiman, and L.
  Schimansky-Geier, Phys. Report 392, 321 (2004).
\bibitem{shulgin}B. Shulgin, A. Neiman, and V. Anishchenko Phys. Rev.
  Lett. 75, 4157 (1995).
\bibitem{longtin2} A. Longtin and D. R. Chialvo, Phys. Rev. Lett. 81,
  4012-4015 (1998).
\bibitem{zhou} Changsong Zhou, J. Kurths, and Bambi Hu, Phys. Rev. E
  67, 030101(R) (2003).
\bibitem{laser}J. A. Freund, S. Barbay, S. Lepri, A. Zavatta, G.
  Giacomelli, Fluct. Noise Lett.  3, L195 (2003).
\bibitem{freund} J. A. Freund, A. Neiman and L. Schimansky-Geier ,
  Europhys. Lett. 50, 8 (2000).
\bibitem{freund2} J. A. Freund and L. Schimansky-Geier, Phys. Rev. E 60,
  1304 (1999).
\bibitem{lai} K. Park, Y. Lai, Z. Liu, A. Nachman, Phys. Lett. A
  326, 391 (2004).
\bibitem{wiesenfeld} B. McNamara, K. Wiesenfeld, Phys. Rev. A 39, 4854
  (1989).
\bibitem{loefstedt} R. L\"ofstedt, S.N. Coppersmith, Phys. Rev. E 49,
  4821 (1994).
\bibitem{talkner3} P. Talkner and J. \L uczka, Phys. Rev. E 69, 046109
  (2004).
\bibitem{longtin} A. Longtin, in {\em Proceedings of the NATO ARW
    Stochastic Resonance in Physics \& Biology}, edited by F. Moss, A.
  Bulsara, and M. F. Shlesinger, J. Stat.  Phys. 70, 309 (1993).
\bibitem{wiesenfeld2} K. Wiesenfeld, D. Pierson, E. Pantazelou, C.
  Dames, and F. Moss, Phys. Rev. Lett. 72, 2125 (1994).
\bibitem{lindner} B. Lindner, L. Schimansky-Geier, Phys. Rev E 61, 6103, 2000
\bibitem{prager1} T. Prager, B. Naundorf, L. Schimansky-Geier, Physica
  A 325, 176 (2003).
\bibitem{prager2} T. Prager, L. Schimansky-Geier, Phys. Rev. Lett. 91,
  230601 (2003).
\bibitem{cox} D. R. Cox, {\em Renewal Theory}, Methuen, London (1962).
\bibitem{harms} T. Harms, R. Lipowsky, Phys. Rev. Letters 79, 2895,
  (1997).
\bibitem{talkner2} P. Talkner, L. Machura, M. Schindler, P. Hänggi,
  J. \L uczka, New J. Phys., submitted.
\bibitem{casado2}J. Casado-Pascual, J. G\'omez-Ord{\'o}\~nez, M. Morillo,
  J. Lehmann, I. Goychuk, P. Hänggi, 
arXiv:cond-mat/0410086
\bibitem{gibson}M. A. Gibson, J. Bruck, J. Phys. Chem. A 104, 1876
  (2000).
\bibitem{comment}Eqs.(\ref{eqp1excitable}) and (\ref{eqp2excitable}) 
 however in contrast to eqs.  (\ref{eqp1excitableasymp}) and
  (\ref{eqp2excitableasymp}) have no unique periodic
  solution even if  supplemented with the
  normalization condition $p_1(t)+p_2(t)=1$.
\bibitem{parmananda} P. Parmananda, C. H. Mena, and G. Baier,
  Phys. Rev. E 66, 047202 (2002)
\bibitem{lee}S.-G. Lee and S. Kim, Phys. Rev. E 60, 826 (1999)
\bibitem{fhn1} R. FitzHugh, Biophys. J. 1, 445 (1961); Biological
  Engineering, McGraw-Hill (1969).
\bibitem{fhn2} J. Nagumo, S. Arimoto, S. Yoshitzawa, Proc. IRE 50,
  2061 (1962).
\end{thebibliography}
\end{document}